# Reconstruction mechanism with self-growing equations for hyper, improper and proper fractional chaotic systems through a novel non-Lyapunov approach[☆]

GAO Fei[a,b,*], FEI Feng-xia[a], XU Qian[a], DENG Yan-fang[a], QI Yi-bo[a], Ilangko Balasingham[b]

[a]*Department of Mathematics, School of Science, Wuhan University of Technology, Luoshi Road 122, Wuhan, Hubei,430070, People's Republic of China*
[b]*Signal Processing Group, Department of Electronics and Telecommunications,Norwegian University of Science and Technology, N-7491 Trondheim, Norway*

## Abstract

Identification of the unknown parameters and orders of fractional chaotic systems is of vital significance in controlling and synchronization of fractional-order chaotic systems. However there exist basic hypotheses in traditional

[☆]The work was carried out during the tenure of the ERCIM Alain Bensoussan Fellowship Programme, which is supported by the Marie Curie Co-funding of Regional, National and International Programmes (COFUND) of the European Commission. The work is also supported by Supported by Scientific Research Foundation for Returned Scholars from Ministry of Education of China (No. 20111j0032), the HOME Program No. 11044(Help Our Motherland through Elite Intellectual Resources from Overseas) funded by China Association for Science and Technology, the NSFC projects No. 10647141, No.60773210 of China, the Natural Science Foundation No.2009CBD213 of Hubei Province of China, the Fundamental Research Funds for the Central Universities of China, the self–determined and innovative research funds of WUT No. 2012–Ia–035, 2012-Ia-041, 2010–Ia–004, The National Soft Science Research Program 2009GXS1D012 of China, the National Science Foundation for Post–doctoral Scientists of China No. 20080431004.

[*]Corresponding author
  *Email addresses:* hgaofei@gmail.com (GAO Fei ), 1092285218@qq.com (FEI Feng-xia), 121214020@qq.com (XU Qian), 986627833@qq.com (DENG Yan-fang), manshengyibo@163.com (QI Yi-bo), ilangko.balasingham@iet.ntnu.no (Ilangko Balasingham)
  *URL:* http://feigao.weebly.com (GAO Fei )

estimation methods, that is, the parameters and fractional orders are partially known or the known data series coincide with definite forms of fractional chaotic differential equations except some uncertain parameters and fractional orders. What should I do when these hypotheses do not exist?

In this paper, a non-Lyapunov novel approach with a novel united mathematical model is proposed to reconstruct fractional chaotic systems, through the fractional-order differential equations self-growing mechanism by some genetic operations ideas independent of these hypotheses. And the cases of identifying the unknown parameters and fractional orders of fractional chaotic systems can be thought as special cases of the proposed united mathematical reconstruction method in non-Lyapunov way. The problems of fractional-order chaos reconstruction are converted into a multiple modal non-negative special functions' minimization through a proper translation, which takes fractional-order differential equations as its particular independent variables instead of the unknown parameters and fractional orders. And the objective is to find best form of fractional-order differential equations such that the objective function is minimized. Simulations are done to reconstruct a series of hyper and normal fractional chaotic systems. The experiments' results show that the proposed self-growing mechanism of fractional-order differential equations with genetic operations is a successful methods for fractional-order chaotic systems' reconstruction, with the advantages of high precision and robustness.

---

1. **Introduction**

Although the concept of fractional calculus was proposed by Leibniz 300 years ago[1–5], it become a powerful tool only during the past decades to describe the dynamics of complex systems in area of science and engineering[6–11].

The applications of fractional differential equations began to appeal to related scientists[12–37] after the symbolic event that Mandelbort discovered that there were a lot of fractional dimension phenomena in nature[38] in following areas, bifurcation, hyperchaos, proper and improper fractional-order chaos systems and chaos synchronization[6–37]. And the integer order chaos systems are the special cases of fractional order chaos systems. It is much more important to research the hyper fractional-order chaos systems, both proper and improper fractional orders.

For the fractional-order chaos systems, most of the control and synchronization methods are generalized conclusive but more complicated approaches from the methods for normal and hyper chaos system. And there are no results on hyper or improper fractional-order chaos systems up to now.

However, there are some systematic parameters and orders are unknown for the fractional-order chaos systems in controlling and synchronization. It is difficult to identify the parameters in the fractional-order chaotic systems with unknown parameters.

Hitherto, there have been at two main approaches in parameters' identi-fication for fractional-order chaos systems.

The first is a Lyapunov way using synchronisation. And there have been few results on parameter estimation method of fractional-order chaotic systems based on chaos synchronization[39]. However, the design of controller and the updating law of parameter identification is a task with technique and sensitively depends on the considered systems.

The second is a non-Lyapunov way using some artificial intelligence methods, such as differential evolution[17] and particle swarm optimization[19]. Although there are many methods in parameters estimation for integer order chaos systems, to the best of our knowledge, a little work in non-Lyapunov way has been done to the parameters and orders estimation of fractional-order chaos systems[6, 17, 19].

For example, we consider the following fractional-order chaos system.

$$_\alpha D_t^q Y(t) = f(Y(t), Y_0(t), \theta) \tag{1}$$

where $Y(t) = (y_1(t), y_2(t), ..., y_n(t))^T \in \Re^n$ denotes the state vector. $\theta = (\theta_1, \theta_2, ..., \theta_n)^T$ denotes the original parameters. $q = (q_1, q_2, ..., q_n), (0 < q_i < 1, i = 1, 2, ..., n)$ is the fractional derivative orders.

$$f(Y(t), Y_0(t), \theta) = (f_1, f_2, ..., f_n)|_{(Y(t), Y_0(t), \theta)}$$

Normally the function $f$ is known. And the $\theta, q$ are unknown , then the $\Theta = (\theta_1, \theta_2, ..., \theta_n, q_1, q_2, ..., q_n)$ will be the parameters to be estimated.

Then a correspondent system are constructed as following.

$$_\alpha D_t^{\tilde{q}} \tilde{Y}(t) = f(\tilde{Y}(t), Y_0(t), \tilde{\theta}) \tag{2}$$

where $\tilde{Y}(t), \tilde{\theta}, \tilde{q}$ are the correspondent variables to those in equation (1), and function $f$ are the same. The two systems (1) (2) have the same initial condition $Y_0(t)$.

And the objective function to be minimized by artificial intelligence methods can be defined as following.

$$F = \sum_{i=1}^{N} \left\| Y_i - \tilde{Y}_i \right\|^2 \tag{3}$$

$$G = \frac{1}{N} \sum_{i=1}^{N} \left\| Y_i - \tilde{Y}_i \right\|^2 \tag{4}$$

$$H = \sum_{i=1}^{N} \left\| Y_i - \tilde{Y}_i \right\|_2 \tag{5}$$

$$W = \frac{1}{N} \sum_{i=1}^{N} \left\| Y_i - \tilde{Y}_i \right\|_2 \tag{6}$$

Then the objective is obtained as following,

$$\theta^* = \arg\min_{\theta} F \tag{7}$$

where $F$ in equation (7) can be any one of equations (3),(4),(5), (6).

However, there exist basic hypotheses in traditional non-Lyapunov estimation methods[6, 17, 19]. That is, the parameters and fractional orders are partially known or the known data series coincide with definite forms $f = (f_1, f_2, ..., f_n)$ of fractional chaotic differential equations except some uncertain parameters and fractional orders $\Theta = (\theta_1, \theta_2, ..., \theta_n, q_1, q_2, ..., q_n)$.

What should I do when these hypotheses do not exist? That is, when some the fractional chaotic differential equations $f = (f_1, f_2, ..., f_n)$ are unknown, how to identify the fractional system? That is,

$$(f_1, f_2, ..., f_n)^* = \arg \min_{(f_1, f_2, ..., f_n)} F \qquad (8)$$

Now the problem of parameters estimation (7) become another much more complicated question, to find the forms of fractional order equations as in (8). In another word, it is fractional-order chaos reconstruction problem now.

However, to the best of authors' knowledge, there are no methods in non-Lyapunov way for fractional order chaotic systems' reconstruction so far. The objective of this work is to present a novel simple but effective approach to reconstruct hyper, proper and improper fractional chaotic systems in a non-Lyapunov way. In which, an idea of self growing equations through only genetic operations are used to identify unknown forms of differential equations $f = (f_1, f_2, ..., f_n)$, in which no extra evolutionary algorithms to be added or be compounded. And the illustrative reconstruction simulations in different chaos systems system are discussed respectively.

The rest is organized as follows. Section 2 give a simple review on non-Lyapunove parameters estimation methods for fractional-order chaos systems. In Section 3, a novel united mathematical model for fractional chaos reconstruction are proposed in a non-Lyapunov way. Then all the existing non-Lyapunov methods in Section 2 become special cases of the new united model, such as non-Lyapunov parameters identification for normal and fractional chaos systems, non-lyapunov reconstruction methods for normal chaos system. In section 4 a novel methods with proposed united model in Section 3 is introduced to reconstruct the hyper, improper and normal fractional chaos systems. And simulations are done to a series of different fractional chaos systems and their special cases in three dimension. Conclusions are summarized briefly in Section 5.

## 2. Non-Lyapunove reconstruction methods for normal chaos systems

To solve the question proposed in Section 1, a new ideas are introduced to generalize the normal parameters' estimation problems into fractional order chaos reconstruction problems. In the sense of the differential equations for chaos system reconstruction, the parameters' identification for normal chaos system cases[40–51] and the for the fractional Lorénz system[13, 18, 34][26, 52] can be thought as special cases of fractional order chaos reconstruction, when the exact forms of fractional order chaotic differential equations $f = (f_1, f_2, ..., f_n)$ are available but some parameters unknown.

To propose a novel reconstruction way for fractional-order chaos system, we briefly reviewed the normal chaos reconstruction methods as following. Firstly, we introduce the basic artificial intelligent symbolic regression method Genetic programming (GP) . Secondly, the non-Lyapunov way for normal chaos reconstruction.

*2.0.1. Genetic Programming*

To have a deep understanding of reconstruction method in non-Lyapunov way, firstly we introduce some basic ideas of GP.

GP proposed by J. R. Koza in the early 1990s[53, 54], belongs to evolutionary algorithms, used to automatically generate the complex function between the input and output data. The basic idea of GP is similar to genetic algorithm. However the more complexity of structure developed in GP's self-adaptive "Program" evolution simulation in finding the best forms of the functions. Because GP uses a much more natural and flexible representation, then it has a wide applications in fields of artificial intelligence, machine learning, control, molecular biology and etc.

According to Darwinian principle of natural selection, individuals with high fitness survived by genetic operators, including crossover, mutation, gene duplication, gene deletion, to the next generation of groups. As the evolution continues, the best or the satisfied optimal solution of the given problem will appears. There are five major preparatory steps for the basic GP to be specified to be evolved.

- Terminals set. The independent variables, zero-argument functions and random constants.

- Set of primitive functions. For instances, $+, -, \times, \div, \sin, \cos, \log, ...$ combine in term of tree structure or the other structure for the individuals.

- Fitness measure. (Explicit or implicit methods measuring the fitness of individuals. Similar to the genetic algorithm, an evaluation criteria is then defined as individuals's fitness values.

- Controlling parameters for the GP's evolution.

- Termination criterion and method for designating the result.

*2.0.2. Evolutionary Reconstruction methods*

The evolutionary reconstruction methods here can be concluded into two categories.

The first is a compound symbolic regression(SR) method[55] with GA, differential evolutionary algorithms or self-organizing migrating algorithm (SOMA) [56–59] and the other evolutionary algorithms. However it can be used for reconstruction of only one differential equation of chaotic systems such as for Lorenz and the other systems[58, 60]. The differences between Ref. [59] and the series researches of Zelinka[56–58] are the fix candidate coding table pre-defined [59] (that is, the basic components of the function form to be searched) and fitness evaluation through Lyapunov exponents.

The second is with model with the parameters unknown in a co-evolution. And the models are to be reconstructed by GP. The optimal parameters' combinations of the reconstructed model are identified by GA, PSO[61–66].

It is also noticed that there is no method for reconstruction for hyperchaos systems and fractional chaos systems so far.

## 3. Unified Novel Mathematical model for fractional chaos reconstruction in non-Lyapunov way

Unlike the normal Lyapunov methods[67–69], for the normal chaos systems, there have been some results on chaos reconstruction for $f = (f_1, f_2, ..., f_n)$ with the non-Lyapunove methods mainly from symbolic regression through genetic programming(GP)[53, 54, 70], and some evolutionary algorithms[56, 58, 60, 62–66, 71–73].

However, to the best of authors' knowledge, there are no direct methods for fractional chaotic systems' reconstruction.

we take the fractional Lorénz system[13, 18, 34] for instance, which is generalized from the first canonical chaotic attractor found in 1963, Lorénz system[74].

$$\begin{cases} {}_\alpha D_t^{q_1} x = \sigma \cdot (y - x); \\ {}_\alpha D_t^{q_2} y = \gamma \cdot x - x \cdot z - y; \\ {}_\alpha D_t^{q_3} z = x \cdot y - b \cdot z. \\ L = (x, y, z) \end{cases} \quad (9)$$

where $q_1, q_2, q_3$ are the fractional orders. When $(q_1, q_2, q_3) = (0.993, 0.993, 0.993)$, $\sigma = 10, \gamma = 28, b = 8/3, \alpha = 0$, intimal point $(0.1, 0.1, 0.1)$ system (9) is chaotic. Generally when the dimension

$$\sum = q_1 + q_2 + q_3 > 2.91$$

for fractional system (9) is chaotic.

The original system is Eq. (9) or matrix data observed from the actual system. Then we generate the three unknown equations $f_i(x, y, z), i = 1, 2, 3$ for reconstructing the chaos system (10).

$$\begin{cases} {}_\alpha D_t^{q_1} \tilde{x} = f_1(\tilde{x}, \tilde{y}, \tilde{z}) = \sigma \cdot (\tilde{y} - \tilde{x}); \\ {}_\alpha D_t^{q_2} \tilde{y} = f_2(\tilde{x}, \tilde{y}, \tilde{z}) = \gamma \cdot \tilde{x} - \tilde{x} \cdot \tilde{z} - \tilde{y}; \\ {}_\alpha D_t^{q_3} \tilde{z} = f_3(\tilde{x}, \tilde{y}, \tilde{z}) = \tilde{x} \cdot \tilde{y} - b \cdot \tilde{z}. \\ \tilde{L} = (\tilde{x}, \tilde{y}, \tilde{z}) \end{cases} \quad (10)$$

Let $f_3(\tilde{x}, \tilde{y}, \tilde{z})$ be unknown in system (10), and the objective here is to reconstruct the $f_3$.

And there exist several definitions of fractional derivatives. Among these, the Grünwald-Letnikov (G-L), Riemann-Liouville (R-L) and the Caputo fractional derivatives are the commonly used[2–5, 75, 76]. And G-L, R-L and Caputo fractional derivatives are equivalent under some conditions[77].

The continuous integro-differential operator[78, 79] is defined as

$$_\alpha D_t^q = \begin{cases} \frac{d^q}{dx^q}, q > 0; \\ 1, q = 0; \\ \int_\alpha^1 (d\tau)^q. \end{cases}$$

We consider the continuous function $f(t)$. The G-L fractional derivatives are defined as following.

$$_\alpha D_t^q f(t) = \lim_{h \to 0} \frac{1}{h^q} \sum_{j=0}^{\left[\frac{t-\alpha}{h}\right]} (-1)^j \binom{q}{j} f(t - jh) \tag{11}$$

where $[x]$ means the integer part of $x$, $\alpha, t$ are the bounds of operation for $_\alpha D_t^q f(t)$, $q \in \Re$ and

$$\binom{q}{j} = \frac{q!}{j!\,(q-j)!} = \frac{\Gamma(q+1)}{\Gamma(j+1)\Gamma(q-j+1)}$$

and $\binom{q}{0} = 1$.

We take ideas of a numerical solution method[78, 79] obtained by the relationship (11) derived from the G-L definition to resolve system (9) and system (10). That is,

$$_{(k-\frac{L_m}{h})} D_{t_k}^q f(t) \approx \frac{1}{h^q} \sum_{j=0}^{k} (-1)^j \binom{q}{j} f(t_{k-j})$$

where $L_m$ is the memory length, $t_k = kh$, $h$ is the time step of calculation and $(-1)^j \binom{q}{j}$ are binomial coefficients $c_j^{(q)}$, $(j = 0, 1, ...,)$. When for numerical computation, the following are used,

$$c_0^{(q)} = 1, c_j^{(q)} = \left(1 - \frac{1+q}{j}\right) c_{j-1}^{(q)}$$

Then in general[78, 79]

$$_\alpha D_t^q f(y(t)) = f(y(t), t)$$

can be expressed as

$$y(t_k) = f(y(t_k), t_k) h^q - \sum_{j=v}^{k} c_j^{(q)} y(t_{k-j})$$

where $v$ in above is defined as

$$v = \begin{cases} k - \frac{L_m}{h}, k > \frac{L_m}{h} \\ 1, k < \frac{L_m}{h} \end{cases}$$

or $v = 1$ for all $k$.

With the ideas above, the method[78, 79] are used here to resolve the fractional Lorénz system (9) with $\alpha = 0$ as following. The general forms of chaos systems (12) and (13) are considered respectively as following, where $i = 1, ..., n$ in equations (12), (13), (14), (15). $(\tilde{f}_1, \tilde{f}_2, ..., \tilde{f}_n)$ in (13) is to be reconstructed.

$$\begin{cases} _\alpha D_t^q x_i = f_i(x_1, x_2, ..., x_n) \\ L = (x_1, x_2, ..., x_n) \end{cases} \quad (12)$$

$$\begin{cases} _\alpha D_t^q \tilde{x}_i = \tilde{f}_i(\tilde{x}_1, \tilde{x}_2, ..., \tilde{x}_n) \\ \tilde{L} = (\tilde{x}_1, \tilde{x}_2, ..., \tilde{x}_n) \end{cases} \quad (13)$$

To have simple forms, we take $\alpha = 0$. With the ideas above, the systems (12) and (13) are solved as following,

$$x_i(t_k) = f_i(x_1(t_k), ..., x_{i-1}(t_k), x_i(t_{k-1}), ..., x_n(t_{k-1})) h^{q_i} - \sum_{j=v}^{k} c_j^{(q_i)} x_i(t_{k-j})$$

(14)

$$\tilde{x}_i(t_k) = f_i(\tilde{x}_1(t_k), ..., \tilde{x}_{i-1}(t_k), \tilde{x}_i(t_{k-1}), ..., \tilde{x}_n(t_{k-1})) h^{q_i} - \sum_{j=v}^{k} c_j^{(q_i)} \tilde{x}_i(t_{k-j}) \tag{15}$$

Then the novel objective function (fitness) equation (16) in this paper come into being from equations (14) and (15) as below.

$$F = \sum_{t=0 \cdot h}^{T \cdot h} \left\| \tilde{L} - L \right\|^2 \tag{16}$$

The objective function (16) to be optimized can also be any kind of equations (3),(4),(5), (6).

It should be noticed here that the independent variables in function (16) are not the parameters but the special variables, for instance, as functions $(\tilde{f}_1, \tilde{f}_2, ..., \tilde{f}_n)$. That is

$$(\tilde{f}_1, \tilde{f}_2, ..., \tilde{f}_n)^* = \arg \min_{(\tilde{f}_1, \tilde{f}_2, ..., \tilde{f}_n)} F \tag{17}$$

For the reconstructions of three dimensional fractional order Lorénz system (9) and correspond (10), the objective function can be as following.

$$f_3^* = \arg \min_{f_3} F \tag{18}$$

Equation (17) and (18) is the crucial turning point that changing from the parameters estimation into functions reconstruction, in other words, fractional order chaos systems' reconstruction.

Now, with the definitions of equations (16) and (17), it can be concluded that the parameters' estimation of fractional order chaos system is a special case of fractional order chaos reconstruction. And further, the parameters estimation and reconstruction in Section 2 for the normal chaos system is the special cases of fractional order chaos systems' reconstruction (17).

## 4. A novel self-growing mechanism of fractional chaos systems' reconstruction in non-Lyapunov way

The task of this section is to find a simple but effective approach for hyper, proper and improper fractional-order chaos reconstruction.

### 4.1. A novel approach for for hyper, proper and improper fractional chaos reconstruction

Now we can propose a self-growing differential equations approach for hyper, proper and improper fractional chaos systems. The pseudo-code of the proposed reconstruction is given below.

### 4.2. Hyper, proper and improper fractional chaos

To test the Algorithm 1, some different well known and widely used Hyper, proper and improper fractional chaos systems are choose as following.

Example. 1. Fractional order Arneodo's System (19)[78, 80].

$$\begin{cases} {}_0D_t^{q_1} x(t) = y(t); \\ {}_0D_t^{q_2} y(t) = z(t); \\ {}_0D_t^{q_3} z(t) = -\beta_1 x(t) - \beta_2 y(t) - \beta_3 z(t) + \beta_4 x^3(t). \end{cases} \quad (19)$$

when $(\beta_1, \beta_2, \beta_3, \beta_4) = (-5.5, 3.5, 0.8, -1.0)$, $(q_1, q_2, q_3) = (0.97, 0.97, 0.96)$, initial point $(-0.2, 0.5, 0.2)$, Arneodo's System (19) is chaotic.

Example. 2. Fractional order Duffing's system (20)[78].

$$\begin{cases} {}_0D_t^{q_1} x(t) = y(t); \\ {}_0D_t^{q_2} y(t) = x(t) - x^3(t) - \alpha y(t) + \delta \cos(\omega t). \end{cases} \quad (20)$$

**Algorithm 1** Fractional order chaos system reconstruction with genetic operations

1: **Basic parameters' setting** the basic set $\{\times, \div, +, -\}$ etc., and the input variables $\{x_1, x_2, ..., x_n\}$ and etc.
2: **Initialize** Generate the initial population with tree structures.
3: **while** Termination condition is not satisfied **do**
4:    **Evaluation** Evaluate the fitness with Eq. (17), where system (13) are reconstructed by the individuals.   ▷ fractional order differential equations
5:    **Crossover**
6:    **Mutation**
7:    **Reproduction** Tournament size. ▷ The Elitism rate is $0.02 - 0.05$.
8: **end while**
9: **Output** Global optimum $(f_1, f_2, ..., f_n)$

when $(a, b, c) = (0.15, 0.3, 1)$, $(q_1, q_2) = (0.9, 1)$, initial point $(0.21, 0.31)$, Duffings system (20) is chaotic.

Example. 3. Fractional order Lotka-Volterra system (21)[78].

$$\begin{cases} {}_0D_t^{q_1}x(t) = \alpha x(t) - bx(t)y(t) + ex^2(t) - sz(t)x^2(t); \\ {}_0D_t^{q_2}y(t) = -cy(t) + dx(t)y(t); \\ {}_0D_t^{q_3}z(t) = -pz(t) + sz(t)x^2(t). \end{cases} \quad (21)$$

when $(a, b, c, d, e, p, s) = (1, 1, 1, 1, 2, 3, 2.7)$, $(q_1, q_2, q_3) = (0.95, 0.95, 0.95)$, initial point $(1, 1.4, 1)$, Lotka-Volterra system (21) is chaotic.

Example. 4. Fractional order Genesio-Tesi's System (22)[78, 81].

$$\begin{cases} {}_0D_t^{q_1}x(t) = y(t); \\ {}_0D_t^{q_2}y(t) = z(t); \\ {}_0D_t^{q_3}z(t) = -\beta_1 x(t) - \beta_2 y(t) - \beta_3 z(t) + \beta_4 x^2(t). \end{cases} \quad (22)$$

when $(\beta_1, \beta_2, \beta_3, \beta_4) = (1.1, 1.1, 0.45, 1.0)$, $(q_1, q_2, q_3) = (1, 1, 0.95)$, initial point $(-0.1, 0.5, 0.2)$, Genesio-Tesi's System (22) is chaotic.

Example. 5. Fractional order Lorénz system (9) with the exact form of the differential equation ${}_0D_t^{q_3}z = f_3(x, y, z)$ are unknown. when $(a, b, c) = (10, 28, 8/3)$, $(q_1, q_2, q_3) = (0.993, 0.993, 0.993)$, initial point $(0.1, 0.1, 0.1)$, Lorénz system (9) is chaotic.

Example. 6. Fractional order Lü system (23)[26, 78].

$$\begin{cases} {}_0D_t^{q_1}x(t) = a(y(t) - x(t)); \\ {}_0D_t^{q_2}y(t) = -x(t)z(t) + cy(t); \\ {}_0D_t^{q_3}z(t) = x(t)y(t) - bz(t). \end{cases} \quad (23)$$

when $(a, b, c) = (36, 3, 20)$, $(q_1, q_2, q_3) = (0.985, 0.99, 0.98)$, initial point $(0.2, 0.5, 0.3)$, Lü system (23) is chaotic.

Example. 7. Fractional order Liu System (24)[36, 78, 82].

$$\begin{cases} {}_0D_t^{q_1}x(t) = -ax(t) - ey^2(t); \\ {}_0D_t^{q_2}y(t) = by(t) - kx(t)z(t); \\ {}_0D_t^{q_3}z(t) = -cz(t) + mx(t)y(t). \end{cases} \quad (24)$$

when $(a, b, c, e, k, m) = (1, 2.5, 5, 1, 4, 4)$, $(q_1, q_2, q_3) = (0.95, 0.95, 0.95)$, initial point $(0.2, 0, 0.5)$, Liu System (24) is chaotic.

Example. 8. Fractional order Chen system (25)[37, 78, 83].

$$\begin{cases} {}_0D_t^{q_1}x(t) = a(y(t) - x(t)); \\ {}_0D_t^{q_2}y(t) = (d)x(t) - x(t)z(t) + cy(t); \\ {}_0D_t^{q_3}z(t) = x(t)y(t) - bz(t). \end{cases} \quad (25)$$

when $(a, b, c, d) = (35, 3, 28, -7)$, $(q_1, q_2, q_3) = (0.9, 0.9, 0.9)$, initial point $(-9, -5, 14)$, Chen system (25) is chaotic. And when when $(a, b, c, d) = (35, 3, 28, -7)$, $(q_1, q_2, q_3) = (1, 1.24, 1.24)$, initial point $(3.123, 1.145, 2.453)$, Chen system (25) is an improper chaotic system[20].

Example. 9. Fractional order Rössler System (26)[22, 78].

$$\begin{cases} {}_0D_t^{q_1}x(t) = -(y(t) + z(t)); \\ {}_0D_t^{q_2}y(t) = x(t) + ay(t); \\ {}_0D_t^{q_3}z(t) = b + z(t)(x(t) - c). \end{cases} \quad (26)$$

when $(a, b, c) = (0.5, 0.2, 10)$, $(q_1, q_2, q_3) = (0.9, 0.85, 0.95)$, initial point $(0.5, 1.5, 0.1)$, Rössler System (26) is chaotic.

Example. 10. Fractional order Chuas oscillator (27)[84].

$$\begin{cases} {}_0D_t^{q_1}x(t) = \alpha(y(t) - x(t) + \zeta x(t) - W(w)x(t)); \\ {}_0D_t^{q_2}y(t) = x(t) - y(t) + z(t); \\ {}_0D_t^{q_3}z(t) = -\beta y(t) - \gamma z(t); \\ {}_0D_t^{q_4}w(t) = x(t); \end{cases} \quad (27)$$

where
$$W(w) = \begin{cases} a : |w| < 1; \\ b : |w| > 1. \end{cases}$$

when $(\alpha, \beta, \gamma, \zeta, a, b) = (10, 13, 0.1, 1.5, 0.3, 0.8)$, $(q_1, q_2, q_3, q_4) = (0.97, 0.97, 0.97, 0.97)$, initial point $(0.8, 0.05, 0.007, 0.6)$, Chua's oscillator (27) is chaotic.

Example. 11. Hyper fractional order Lorénz System (28)[85] .

$$\begin{cases} {}_0D_t^{q_1} x(t) = a(y(t) - x(t)) + w(t); \\ {}_0D_t^{q_2} y(t) = cx(t) - x(t)z(t) - y(t); \\ {}_0D_t^{q_3} z(t) = x(t)y(t) - bz(t); \\ {}_0D_t^{q_4} w(t) = -y(t)z(t) + \gamma w(t); \end{cases} \quad (28)$$

when $(a, b, c, d) = (10, 8/3, 28, -1)$, $(q_1, q_2, q_3, q_4) = (0.96, 0.96, 0.96, 0.96)$, initial point $(0.5, 0.6, 1, 2)$, Hyper fractional order Lorénz System (28) is chaotic.

Example. 12. Hyper fractional order Lü System (29)[86].

$$\begin{cases} {}_0D_t^{q_1} x(t) = a(y(t) - x(t)) + w(t); \\ {}_0D_t^{q_2} y(t) = -x(t)z(t) + cy(t); \\ {}_0D_t^{q_3} z(t) = x(t)y(t) - bz(t); \\ {}_0D_t^{q_4} w(t) = x(t)z(t) + dw(t); \end{cases} \quad (29)$$

when $(a, b, c, d) = (36, 3, 20, 1.3)$, $(q_1, q_2, q_3, q_4) = (0.98, 0.980.98, 0.98)$, initial point $(1, 1, 1, 1)$, Hyper fractional order Lü System (29) is chaotic.

Example. 13. Hyper fractional order Liu System (30)[87].

$$\begin{cases} {}_0D_t^{q_1} x(t) = -ax(t) + by(t)z(t) + z(t); \\ {}_0D_t^{q_2} y(t) = 2.5y(t) - x(t)z(t); \\ {}_0D_t^{q_3} z(t) = x(t)y(t) - cz(t) - 2w(t); \\ {}_0D_t^{q_4} w(t) = -d \cdot x(t). \end{cases} \quad (30)$$

when $(a, b, c, d) = (10, 1, 4, 0.25)$, $(q_1, q_2, q_3, q_4) = (0.9, 0.9, 0.9, 0.9)$, initial point $(2.4, 2.2, 0.8, 0)$, Hyper fractional order Liu System (30) is chaotic.

Example. 14. Hyper fractional order Chen System (31)[88].

$$\begin{cases} {}_0D_t^{q_1} x(t) = -a(y(t) - x(t)) + w(t); \\ {}_0D_t^{q_2} y(t) = dx(t) - x(t)z(t) + cy(t); \\ {}_0D_t^{q_3} z(t) = x(t)y(t) - bz(t); \\ {}_0D_t^{q_4} w(t) = y(t)z(t) + rw(t). \end{cases} \quad (31)$$

when $(a, b, c, d) = (35, 3, 12, 7, 0.5)$, $(q_1, q_2, q_3, q_4) = (0.96, 0.96, 0.96, 0.96)$, initial point $(0.5, 0.6, 1, 2)$, Hyper fractional order Chen System (31) is chaotic.

Example. 15. Hyper fractional order Rössler System (32)[22].

$$\begin{cases} {}_0D_t^{q_1} x(t) = -(y(t) + z(t)); \\ {}_0D_t^{q_2} y(t) = x(t) + ay(t) + w(t); \\ {}_0D_t^{q_3} z(t) = x(t)z(t) + b; \\ {}_0D_t^{q_4} w(t) = -cz(t) + dw(t). \end{cases} \quad (32)$$

when $(a, b, c, d) = (0.32, 3, 0.50.05)$, $(q_1, q_2, q_3, q_4) = (0.95, 0.950.95, 0.95)$, initial point $(-15.5, 9.3, -4, 18.6)$, Hyper fractional order Rössler System (32) is chaotic.

4.3. Simulations

There is one important thing we have to deal with is that the normal ODE methods are not valid for many individuals in evolution of GP. Even they are efficient in some cases, however the time consuming is too much. Here we assign all the values of NAN and INF as 0 for fractional chaos systems' resolving, then the process of getting results by resolving the fractional order system can be more simpler.

For systems to be reconstructed, the parameters of the proposed method are set as following. Population size 100, max generations 200, mutation rate 0.05, crossover rate is 0.85, directed copy rate is 0.02 (the Elitism rate), the tournament selection is implemented(the size of tournament is 3), the maximum depth of tree is 6, the maximum depth of subtrees created by mutation is 3, the maximum number of nodes per tree is infinite, and the input variables for 3 and 4 dimensional fractional order chaos systems are taken as $x, y, z$ and $x, y, z, w$ respectively. Table 1 give the detail setting for each system.

Table 2 shows the simulation results of above fractional order chaotic systems.

The following figures give a illustration how the self growing evolution process works by genetic operations of Algorithm 1. In which, Figures 1,2 ,3, 4, 5, 6,7 ,8, 9, 10 show the simulation evolution results of above fractional order chaotic systems. Figures 11,12 show some of the correspond objective function's evolution process of above fractional order chaotic systems.

From the simulations results of reconstruction above fractional order chaos system, it can be concluded that the proposed method is efficient. And from above figures, it can be concluded that the estimated systems are self growing under the genetic operations of the proposed methods.

## 5. Conclusions

The put method consists of numerical optimization problem with unknown fractional order differential equations to identify the chaotic systems. Simulation results demonstrate the effectiveness and efficiency of the pro-

Table 1: Detail parameters stetting for different systems

| F-O system | Unknown | Basic set | Constant | Probability | Step | No. of samples |
|---|---|---|---|---|---|---|
| Arneodo | $D_t^{q_1}x, D_t^{q_2}y$ | $\{\times, \div, +, -\}$ | $[-5, 5]$ | 0.1 | 0.005 | 200 |
| Duffing | $D_t^{q_1}x$ | $\{\times, +, -\}$ | | | 0.0005 | 500 |
| Lotka-Volterra | $D_t^{q_1}x$ | $\{\times, +, -\}$ | | | 0.01 | 100 |
| GenTesi | $D_t^{q_1}x, D_t^{q_2}y$ | $\{\times, \div, +, -\}$ | $[-5, 5]$ | 0.1 | 0.005 | 200 |
| Improper Chen | $D_t^{q_1}x$ | $\{\times, +, -\}$ | $[-3, 3]$ | 0.3 | 0.005 | 200 |
| Lorénz | $D_t^{q_1}x$ | $\{\times, +, -\}$ | | | 0.01 | 50 |
| Lorénz | $D_t^{q_3}z$ | $\{\times, +, -\}$ | $[-3, 3]$ | 0.5 | 0.01 | 50 |
| Lü | $D_t^{q_3}z$ | $\{\times, +, -\}$ | $[-5, 5]$ | 0.2 | 0.01 | 100 |
| Liu | $D_t^{q_1}x$ | $\{\times, +, -\}$ | | | 0.01 | 100 |
| Chen | $D_t^{q_3}z$ | $\{\times, +, -\}$ | $[-5, 5]$ | 0.2 | 0.01 | 100 |
| Rössler | $D_t^{q_3}z$ | $\{\times, +, -\}$ | | | 0.01 | 100 |
| ChuaM | $D_t^{q_2}y$ | $\{\times, +, -\}$ | | | 0.01 | 100 |
| Hyper Lorénz | $D_t^{q_4}w$ | $\{\times, +, -\}$ | | | 0.01 | 100 |
| Hyper Lü | $D_t^{q_3}z$ | $\{\times, +, -\}$ | | | 0.005 | 200 |
| Hyper Liu | $D_t^{q_4}w$ | $\{\times, +, -\}$ | $[-1, 1]$ | 0.5 | 0.005 | 100 |
| Hyper Chen | $D_t^{q_3}z$ | $\{\times, +, -\}$ | | | 0.005 | 200 |
| Hyper Rössler | $D_t^{q_3}z$ | $\{\times, +, -\}$ | $[-4, 4]$ | 0.5 | 0.005 | 200 |

Table 2: Simulation results for different fractional order chaos systems

| F-O system | Unknown | StD | Mean[a] | Min | Max | Success rate[b] |
|---:|---:|---|---|---|---|---|
| Arneodo | $D_t^{q_1}x, D_t^{q_2}y$ | 5.2631e-33 | 5.3645e-34 | 0 | 5.2635e-32 | 100% |
| Duffing | $D_t^{q_1}x$ | 0 | 0 | 0 | 0 | 100% |
| Lotka-Volterra | $D_t^{q_1}x$ | 1.9018e-03 | 6.1732e-04 | 0 | 1.6809e-02 | 58% |
| GenTesi | $D_t^{q_1}x, D_t^{q_2}y$ | 0 | 0 | 0 | 0 | 100% |
| Improper Chen | $D_t^{q_1}x$ | 1.9018e-03 | 6.1732e-04 | 0 | 1.6809e-02 | 71% |
| Lorénz | $D_t^{q_1}x$ | 276.7566 | 202.3219 | 0 | 1077.2092 | 37% |
| Lorénz | $D_t^{q_3}z$ | 368.5500 | 38.4626 | 3.3899e-5 | 2367.7693 | 26% [c] |
| Lü | $D_t^{q_3}z$ | 1091.1466 | 264.2664 | 0 | 9922.8725 | 88% |
| Liu | $D_t^{q_1}x$ | 2.1666e-4 | 3.9431e-5 | 0 | 1.6923e-3 | 91% |
| Chen | $D_t^{q_3}z$ | 483.5588 | 146.9202 | 0 | 2800.4232 | 70% |
| Rössler | $D_t^{q_3}z$ | 5.7192e-7 | 7.7360e-8 | 0 | 5.5227e-6 | 97% |
| ChuaM | $D_t^{q_2}y$ | 0.5669 | 0.1353 | 0 | 4.4050 | 85% |
| Hyper Lorénz | $D_t^{q_4}w$ | 2746.4795 | 928.6374 | 0 | 17742.4778 | 79% |
| Hyper Lü | $D_t^{q_3}z$ | 1263.4313 | 251.3628 | 0 | 7222.2843 | 95% |
| Hyper Liu | $D_t^{q_4}w$ | 2.1157e-5 | 5.5160e-6 | 1.0493e-18 | 1.7073e-4 | 56%[d] |
| Hyper Chen | $D_t^{q_3}z$ | 4754.4523 | 1293.1303 | 0 | 30571.5591 | 90% |
| Hyper Rössler | $D_t^{q_3}z$ | 0.8458 | 0.1373 | 4.8299e-14 | 6.0346 | 62% [e] |

[a] Std and Mean are calculated for all the solutions.

[b] Success means the the solution is less than $1e-20$ in 100 independent simulations.

[c] Success means the the solution is less than $1e-1$ in 100 independent simulations.

[d] Success means the the solution is less than $1e-8$ in 100 independent simulations.

[e] Success means the the solution is less than $1e-4$ in 100 independent simulations.

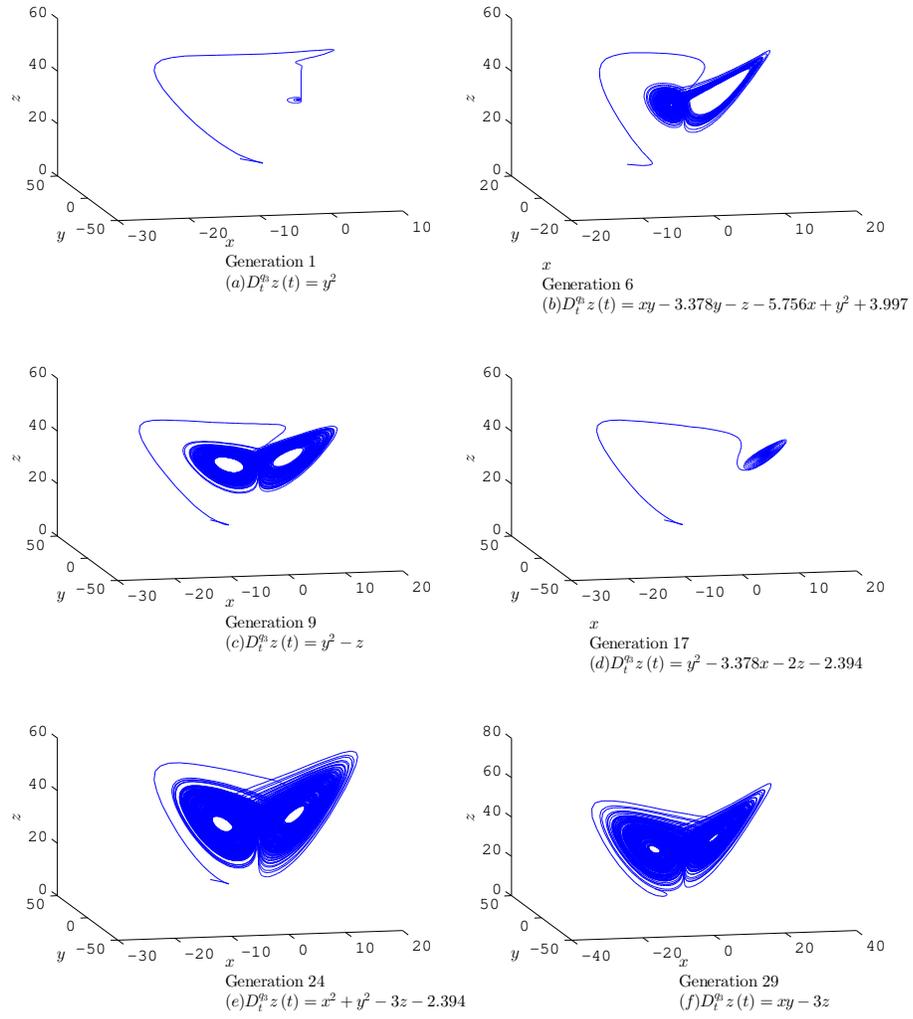

Figure 1: The fractional Chen system with best $D_t^{q_3} z$
23

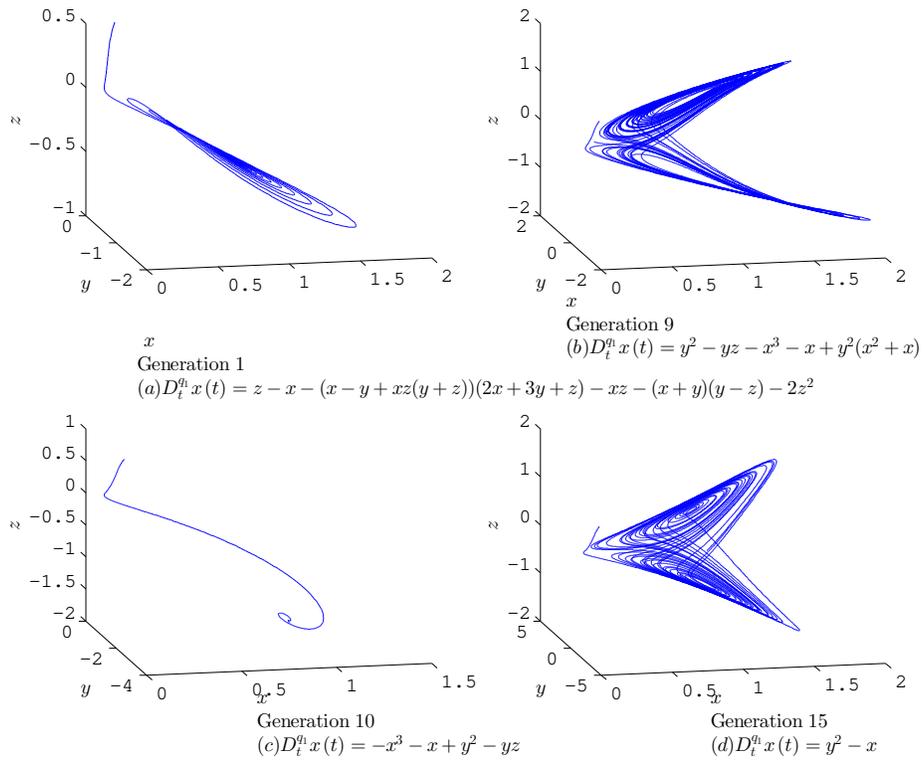

Figure 2: The fractional Liu system with best $D_t^{q_1} x$

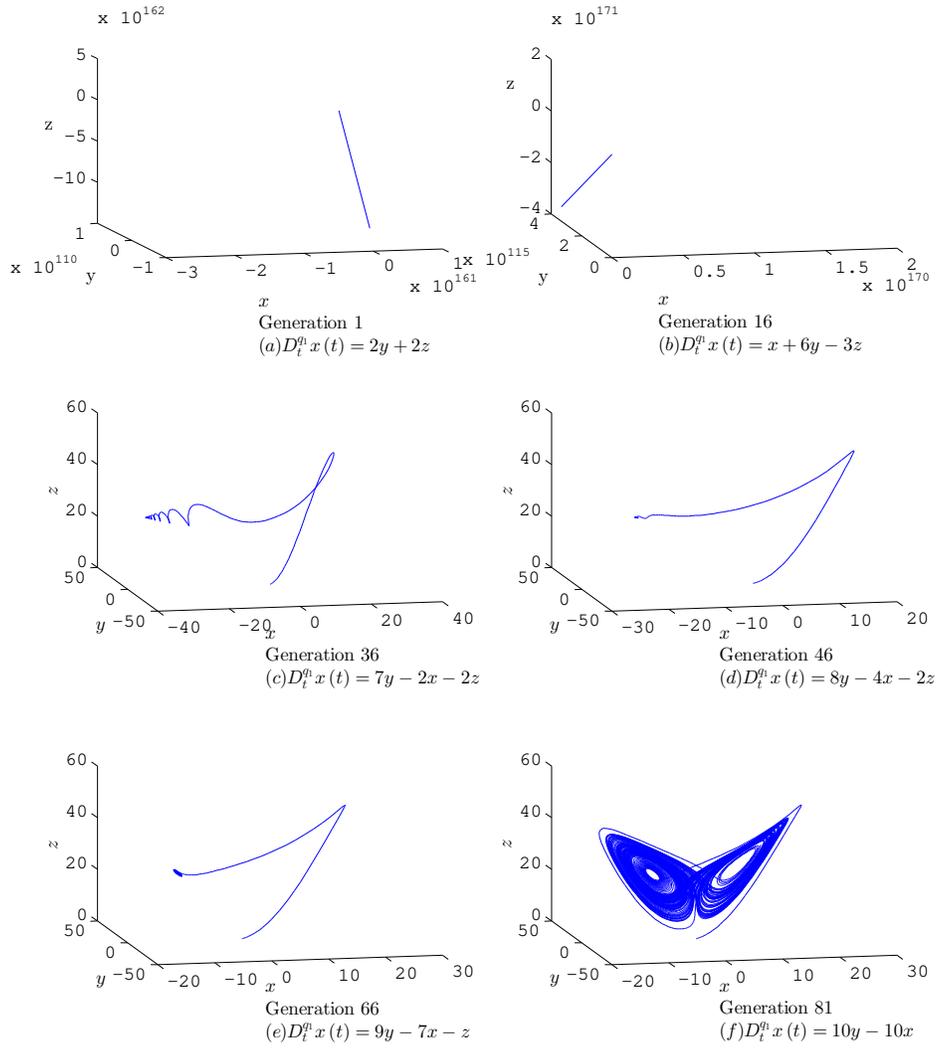

Figure 3: The fractional Lorénz system with best $D_t^{q_1} x$

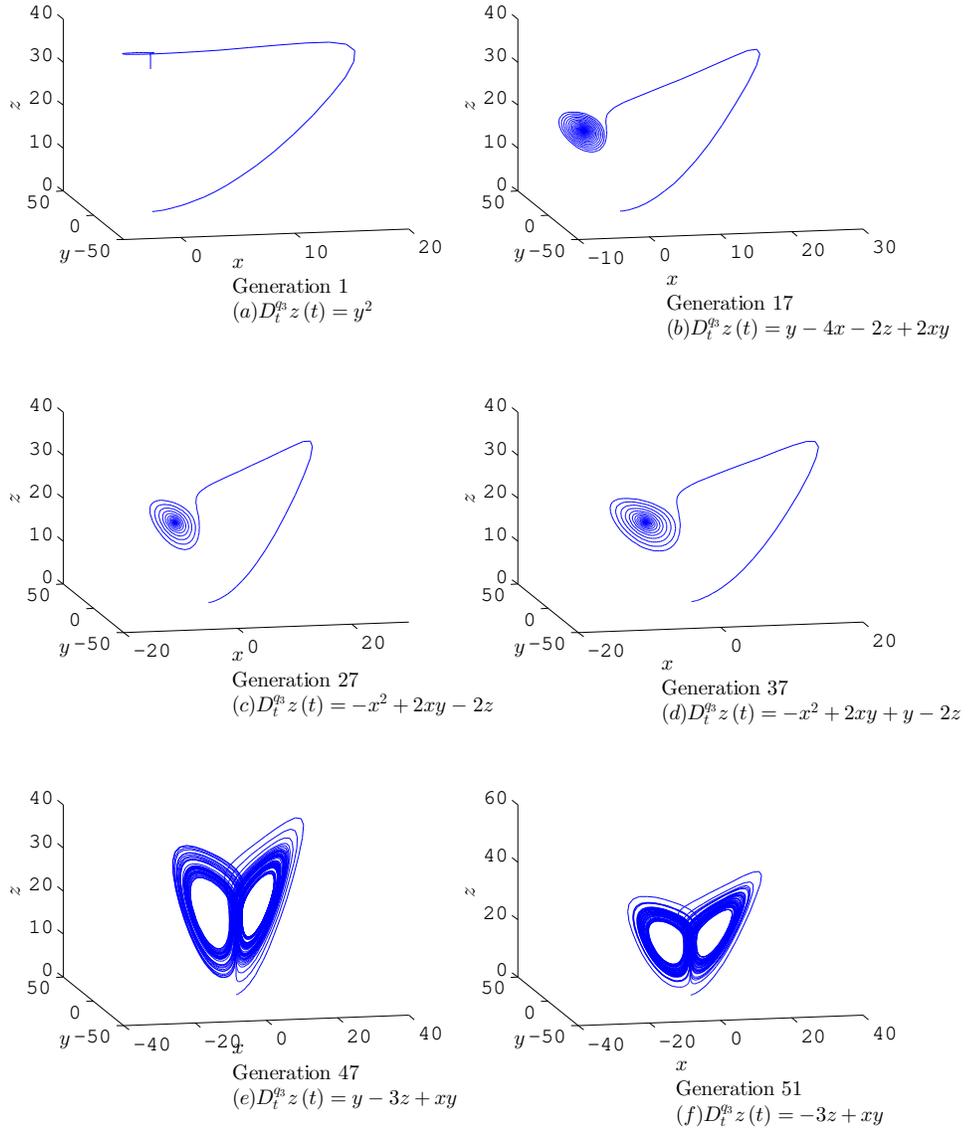

Figure 4: The fractional Lü system with best $D_t^{q_3} z$

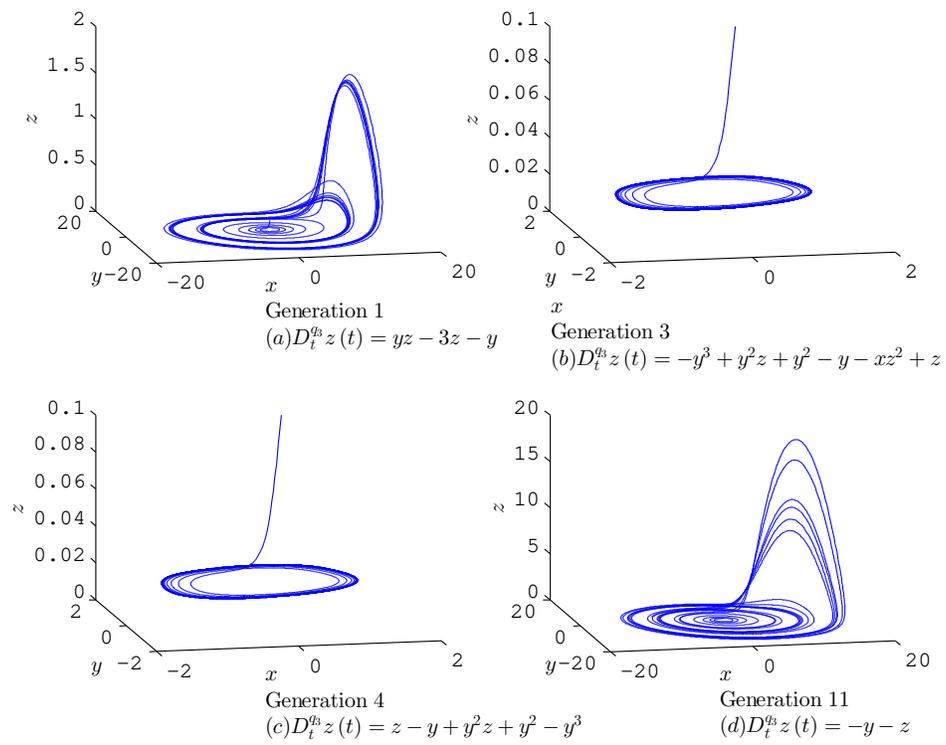

Figure 5: The fractional Rössler system with best $D_t^{q_3} z$

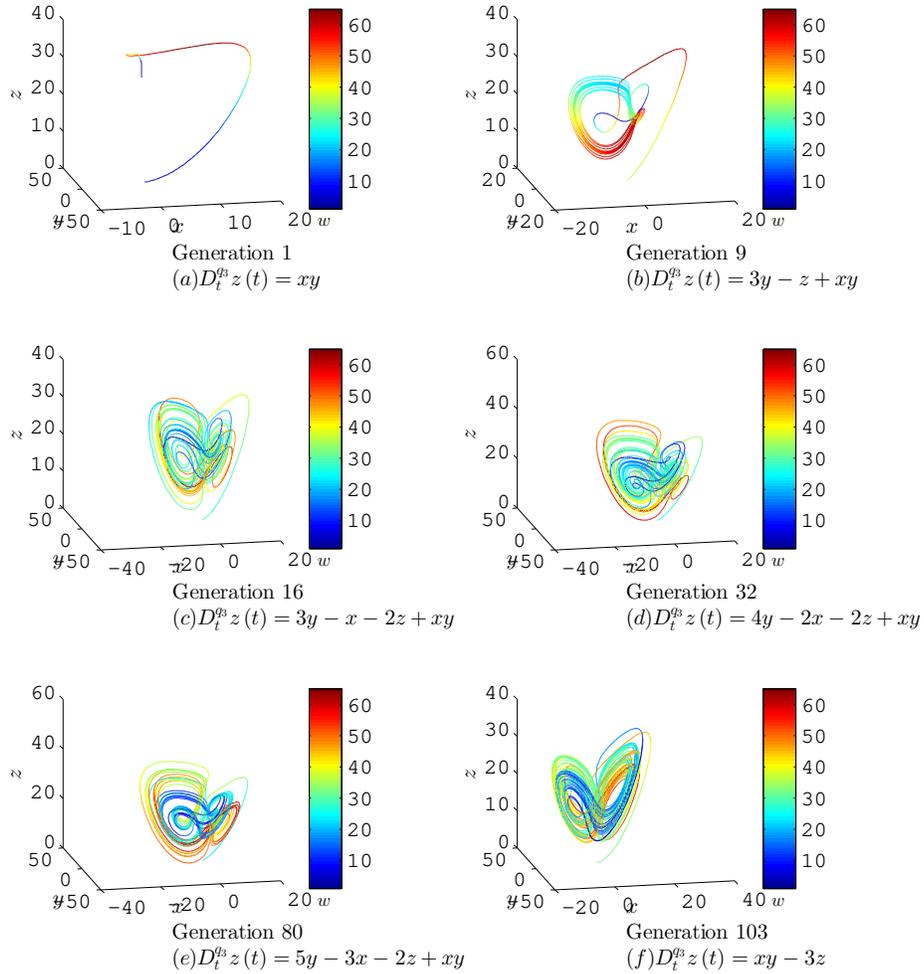

Figure 6: The hyper fractional Chen system with best $D_t^{q_3} z$

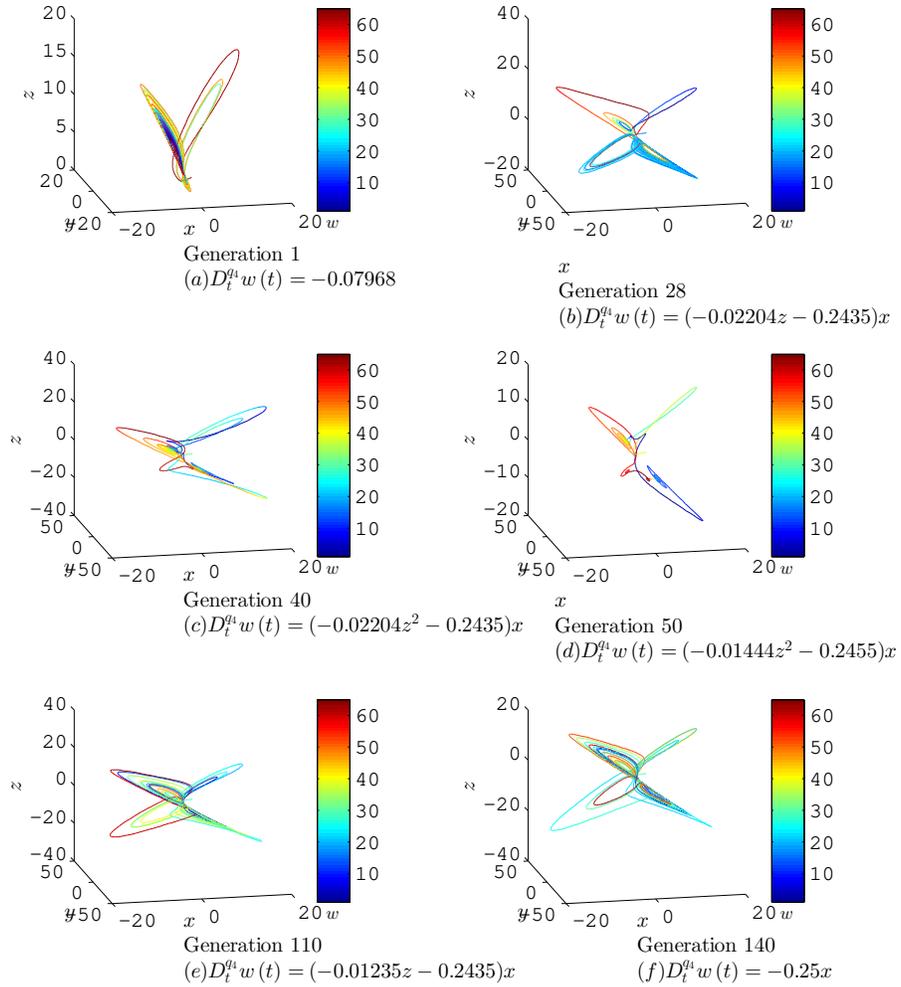

Figure 7: The hyper fractional Liu system with best $D_t^{q_4} w$

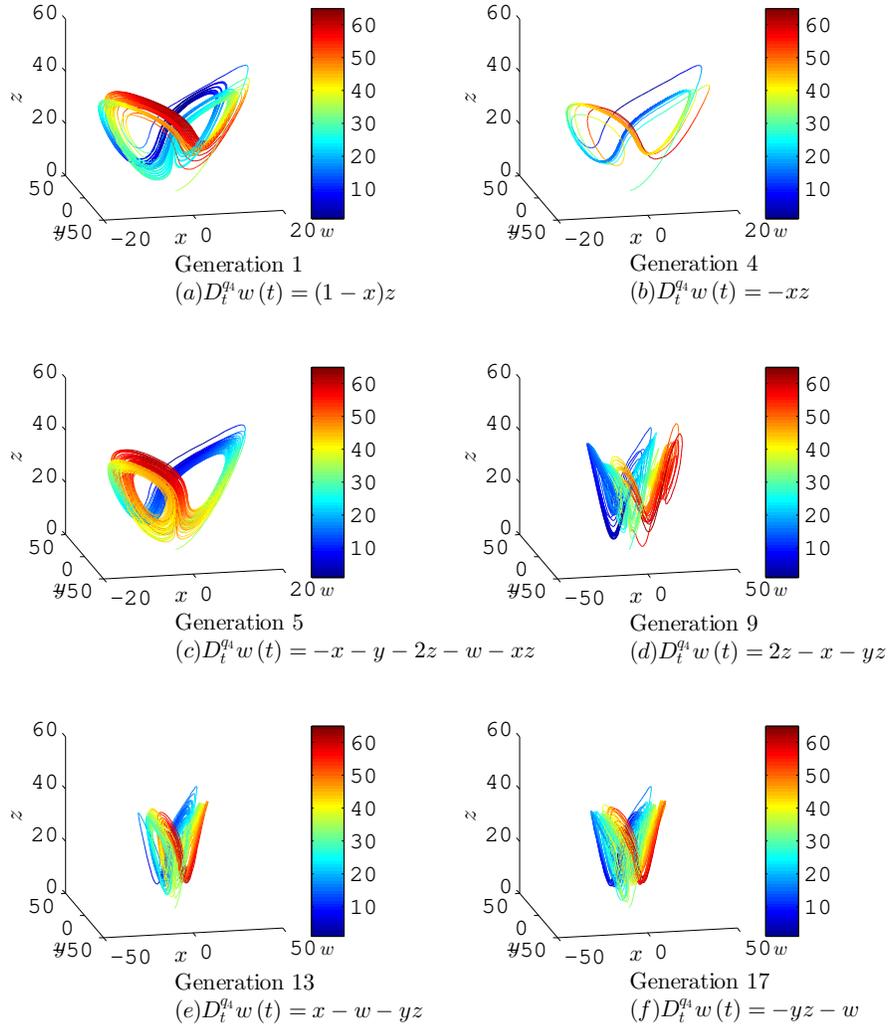

Figure 8: The hyper fractional Lorénz system with best $D_t^{q_4} w$

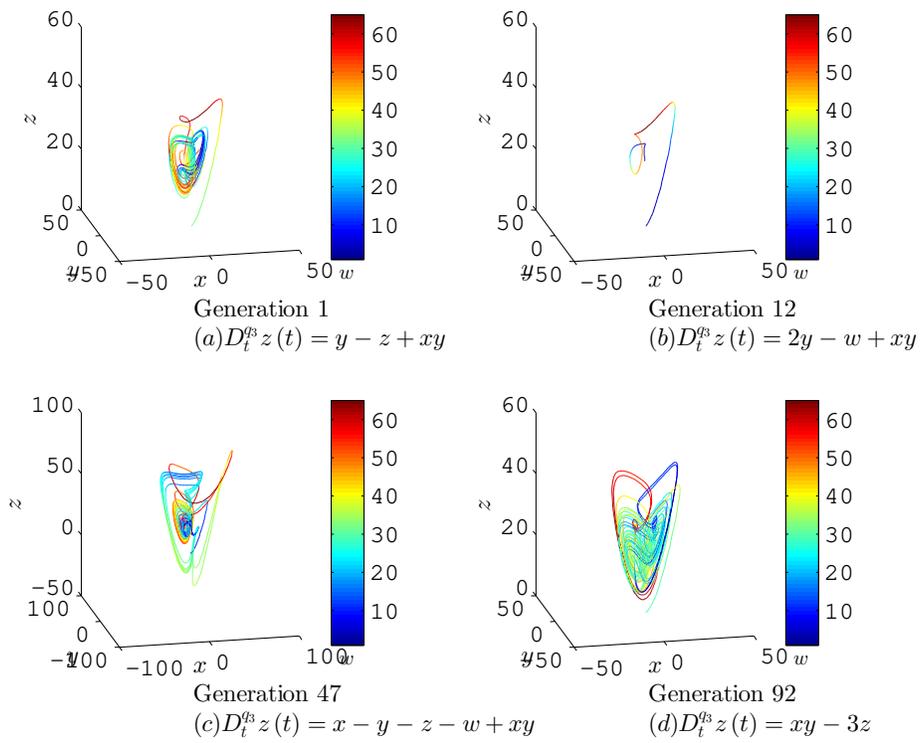

Figure 9: The hyper fractional Lü system with best $D_t^{q_3} z$

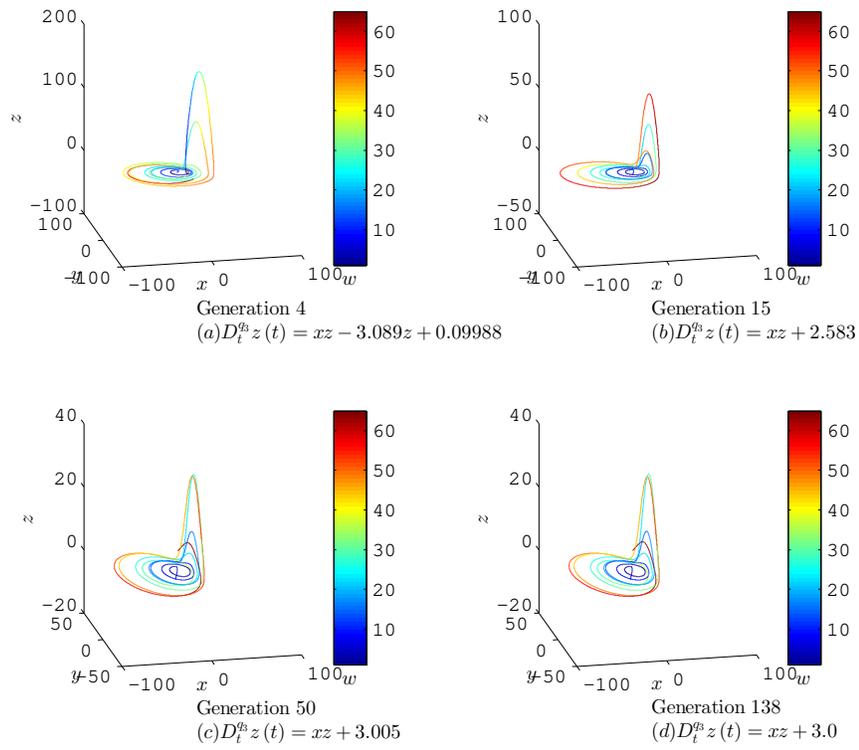

Figure 10: The hyper fractional Rössler system with best $D_t^{q_3} z$

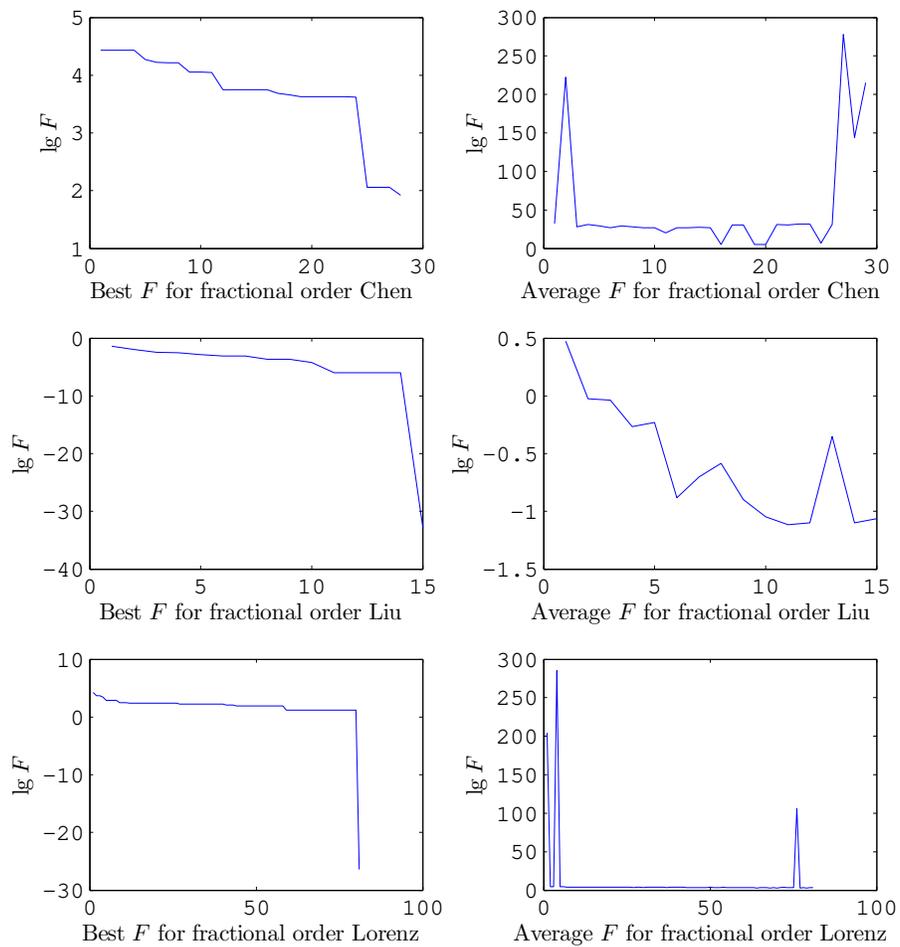

Figure 11: Evolution process of objective function $F$

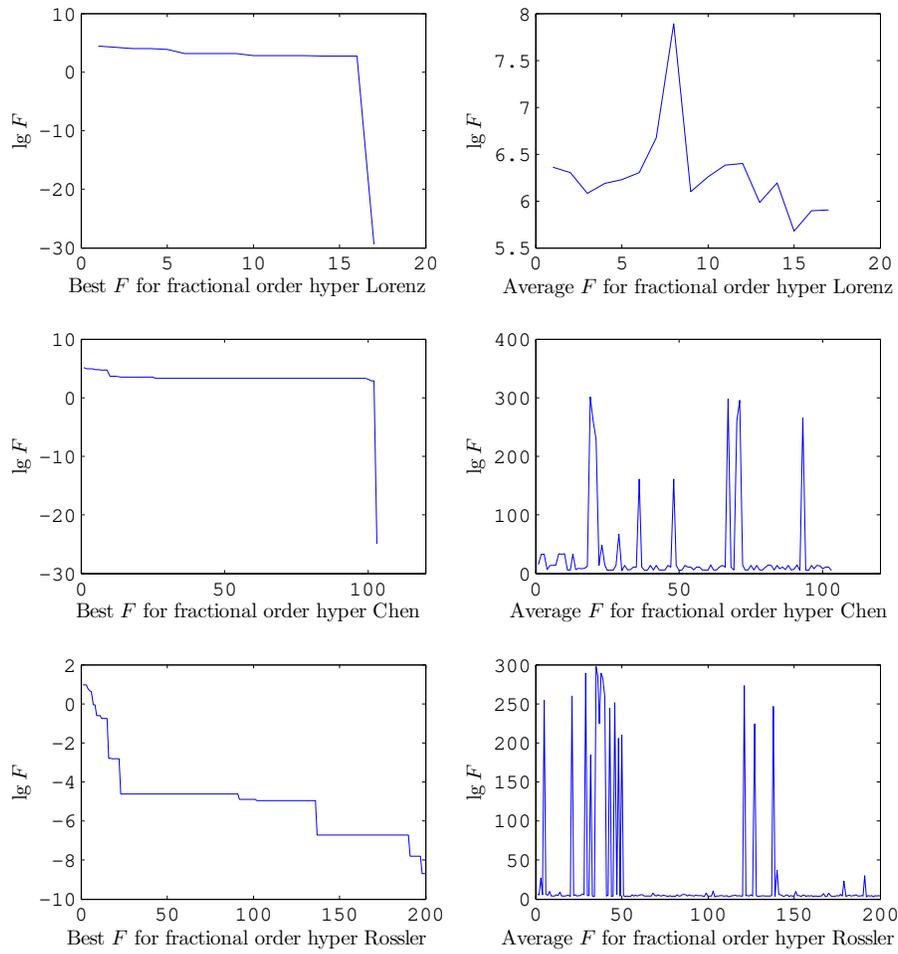

Figure 12: Evolution process of objective function $F$

posed methods with the unified new mathematical model. This is a novel Non–Lyaponov way for chaos reconstruction.

We still have to improve the searching performance for genetic operations of proposed method, for the reason that the more efficient the genetic operations are, the better the new method is in some degree. Moreover, it also illustrated the simplicity and easy implementation of the other kind of intelligent algorithms to replace genetic operations of Algorithms 1 for applications, as in genetic operations from differential evolution algorithms added to tree structures in reference [89].

The performance of the proposed method is sensitive to the initial point for each fractional order chaos system, sample period, number of points, and intervals. Actually, these also lead to the candidate system divergent. And they are not predefined randomly. A good combination of these is not easy to get. Some mathematical formula to get a good combination not by so many simulations will be introduced in the future studies.

It should be noticed too many points for evaluating chaos system the individual represents are not worth. Because the most time consumption parts in the whole proposed method are to resolve the candidate systems. Some of these system are easy to solve. However when it comes with the bad individuals, the methods to resolve the fractional order chaos systems in Section 3 might not converge. Then the whole proposed method Algorithms 1 might get into endless loops. To avoid the endless loops, we introduce a forced strategy to assign all the NAN and infinite numbers in the output as zero. Because the objective function (17) to be optimized is bigger than 0, so this forced strategy for assignment is reasonable. To achieve a fine bal-ance between the performance of the proposed methods and having enough sample data for credibility, we take the number of the points as $100 - 200$, according the existing simulations[40, 41, 43–45, 47, 48, 50, 51, 90–95]. And the simulations in section 4 results show it is effective too.

Here we have to say the success rate has something to do with the problems to be reconstructed themselves. No methods are valid for all the problems. According the success rate of Table 2, there are different results for different fractional chaos systems, even the same methods are used. In which, the worst one is for fraction order Lorenz system and hyper fractional order Rössler and Liu systems. The reason lies in the systems themselves. For $D_t^{q_3} z$ of the fraction order Lorenz system, the best one is $D_t^{q_3} z = x \cdot y - b \cdot z$ with $b = 8/3$. However, it is not easy to achieve $b = 8/3$. As we know from the existing simulations[40, 41, 43–45, 47, 48, 50, 51, 90–95], even the estimated $\tilde{b}$ is very near to real value

$$b = 8/3$$

, the estimated system are not the same as the real Lorénz system. Thus for the Algorithms 1 here, it has two tasks, not only finding the best form for fractional differential equations $D_t^{q_3} z = f_3$, but also finding the best value for $b$. For the hyper fractional order Rössler system, $D_t^{q_3} z = xz + 3.0$ is not easy to be achieved either. Here the key point is the constant $3 \in D_t^{q_3} z = xz + 3.0$. And the same to the hyper fractional order Liu system where $0.25 \in D_t^{q_4} w = 0.25x$

In conclusion, it has to be stated that self growing fractional order differential equations in chaos systems' reconstruction is a promising direction. And the other EAs can also be introduced inside of the proposed methods, by doing their own special genetic operations to the individuals in GP with the objective function (17) .